\documentclass[10pt]{cai26}
\usepackage[most]{tcolorbox}
\newcommand{\ours}{Code-Aware Agent }
\newcommand{\okey}{CA2}

\definecolor{sb_blue}{RGB}{31,119,180}
\begin{document}
\begin{center}

\title{CA2: Code-Aware Agent for Automated Game Testing}
\maketitle

\thispagestyle{empty}
\pagenumbering{gobble}

\begin{tabular}{cc}
Valliappan Chidambaram Adaikkappan\upstairs{\affilone \affilthree}, Vincent Martineau \upstairs{\affiltwo}, Joshua Romoff \upstairs{\affiltwo}, David Meger\upstairs{\affilone \affilthree}
\\[0.25ex]
{\small \upstairs{\affilone} McGill University} \\
{\small \upstairs{\affiltwo} Ubisoft} \\
{\small \upstairs{\affilthree} Mila Quebec AI Institute} \\
\end{tabular}
  
\emails{
  Preprint. Correspondence to: valliappan.chidambaramadaikkappa@mail.mcgill.ca
 
}
\vspace*{0.1in}
\end{center}

\begin{tcolorbox}[
    colback=sb_blue!20,
    colframe=blue!75!black,
    arc=3mm,
    boxrule=0.8pt
]
\vspace{-2pt}
\begin{abstract}

Automated game testing is important for verifying game functionality, but it remains a costly and time-consuming process. Manual testing often misses edge cases, and current automated methods struggle to provide full code coverage. Prior work has explored reinforcement learning (RL) for game testing, but without leveraging internal code signals such as the call stack. We present \ours~(\okey), which uses call stack information to learn effective testing strategies. The agent receives the current function call trace along with the game state and learns to reach specific target functions. We instrument two types of environments, 1) State-based and 2) Image-based, with support for efficient call stack extraction. Through experimental evaluation, we find that \okey~ achieves consistent improvement over the non-code aware baselines, which does not leverage call stack information. Our results show that incorporating code signals like the call stack enables more effective and targeted game testing.

\end{abstract}
\end{tcolorbox}

\section{INTRODUCTION}
Automated game testing is a crucial part of game development, helping developers ensure quality, stability, and performance. However, despite its importance, it remains one of the most tedious and labor-intensive tasks, often requiring large teams of human testers \cite{politowski2021surveyvideogametesting}. Recent trends show that the use of AI tools to aid traditional testing practices, such as manual testing, scripted coverage procedures, and basic unit tests \cite{10333194}. RL agents are potentially excellent candidates as they have been shown to simulate complex human behavior in different video game settings \cite{mnih2013playing, hafner2024masteringdiversedomainsworld}. However, the effectiveness of an automated testing approach that simply replicates human behavior remains largely unexplored. Most prior work in automated game testing focuses on maximizing \emph{state coverage}, ensuring agents visit a wide range of game states~\cite{gordillo2021improvingplaytestingcoveragecuriosity, 9970382, liu2022inspectorpixelbasedautomatedgame}. However, identical game states can trigger diverse function calls based on the chosen action. As a result, state coverage alone fails to guarantee meaningful code coverage. 
\\\\
\begin{figure}[htb!]
    \centering
    \includegraphics[trim={0.75cm 0.50cm 0.75cm 0.3cm}, clip, width=0.6\linewidth]{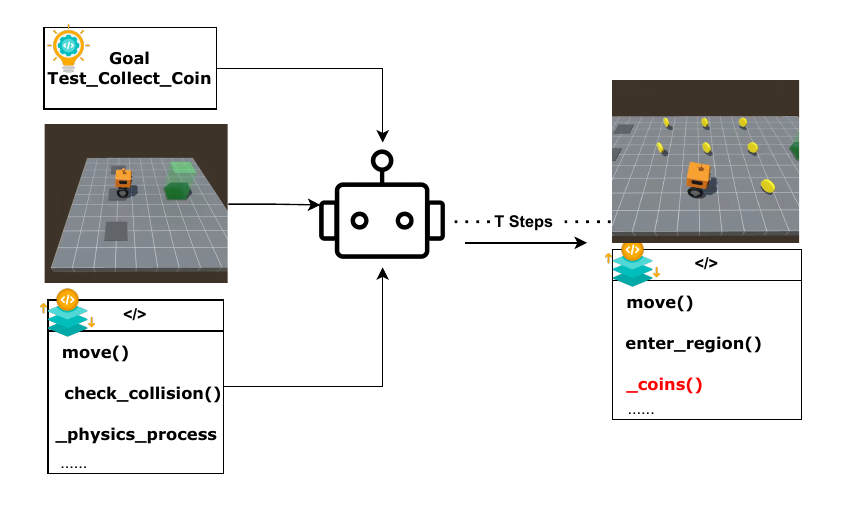}
    \caption{Presenting the \ours~(\okey) for automated game testing. It seamlessly tests low-level game functions by leveraging call stack information.}
    \label{fig:enter-label}
\end{figure}
This paper proposes new research tools and methods relevant for \emph{function-level RL-based game testing}, where agents are trained for a functional test that aims to improve code coverage. Specifically, we present two novel environments designed for functional testing of games, which adhere to the gym environments framework \cite{brockman2016openai}. The first environment builds upon a multilevel Godot game that has been explored in previous RL research \cite{beeching2021godotreinforcementlearningagents}. The second transforms the very well-known Crafter~\cite{hafner2022benchmarkingspectrumagentcapabilities} RL benchmark into a game-testing scenario. Unlike traditional RL environments, we define sparse rewards for reaching functions, which requires the integration of a source code profiler into the traditional gym interface. \\\\
Running real video game engines with a profiler can be slow and difficult to scale for online RL training that requires millions of samples \cite{mnih2013playing}. Thus, we have collected a corpus of expert data gathered by human testers, allowing the use of imitation learning (IL) or offline RL algorithms that do not need to directly interact with the game engine for training.The integration of a source code profiler into our environments allows RL agents to be \textbf{code-aware} while interacting with the game. In addition to typical observations that include images or other transcriptions of the game's simulated physical space, we have experimented with agents that can make use of the call stack as an additional observation. \\\\

To this end, we propose \textit{\ours}, which can understand and test low-level
game-state functions by leveraging call stack information. In our approach, agents play the game and try to reach a targeted method in the source code, thus exploring execution branches, exercising the code base, discovering bugs or crashes, and verifying performance characteristics. We hypothesize that by incorporating programmatic feedback, such as the call stack, RL can outperform traditional code-unaware agents in terms of reaching specific function-level goals. This makes the function-coverage goals observable to the agent, but raises the challenge of multi-modality. The function stack and physical observations have significantly different dimensionalities and unique transition dynamics. We find that naive integration of these state elements does not lead to strong agent performance, but through careful joint representations, our code-aware agents give a boost to testing efficacy. Our approach proves effective across both pixel-based environments with discrete action spaces and state-based environments with continuous actions. \\\\Our main contributions are as follows:
\begin{itemize}
    \item We introduce a suite of two function-level game testing environments, Crafter and Godot, each instrumented with a lightweight source code profiler and equipped with expert gameplay trajectories to support offline reinforcement learning.
    \item A novel code-aware agent design that augments standard offline IL/RL architectures with call stack observations. See Figure~\ref{fig:enter-label}.
    \item The proposed approach is engine-agnostic by design. We demonstrate this through two distinct implementations: Crafter, a pixel-based environment built with the Pygame engine (Python). Multilevel Robot Game, implemented in Godot (GDScript).
    
\end{itemize}

\section{RELATED WORK}
\textbf{Automated Game testing}. Several studies have investigated the use of AI in game testing. \cite{Chen_2021} propose an object detection-based approach for detecting non-crashing game glitches, such as visual glitches across 20 real-world games. This approach relied on a data augmentation approach that can generate game images from User Interface (UI) glitches. Recent advances in large language models (LLMs) have spurred the development of generalist agents for complex video game environments such as StarCraft II\cite{vinyals2019grandmaster}, Minecraft \cite{lifshitz2024steve1generativemodeltexttobehavior}, and Civilization \cite{qi2024civrealmlearningreasoningodyssey}. Early systems often rely on structured APIs or predefined action spaces to interface with games, which simplifies control but limits generalization across environments. End-to-end approaches have demonstrated greater flexibility: VPT\cite{baker2022videopretrainingvptlearning} showed that agents can learn directly from raw video via large-scale pretraining on human gameplay, though such datasets are expensive to collect and difficult to scale. Similarly, SIMA \cite{simateam2024scalinginstructableagentssimulated}trained embodied agents across multiple 3D games using behavior cloning, but remained constrained by high data requirements and limited transfer. More recent work emphasizes autonomous skill acquisition and continual learning. VOYAGER \cite{wang2023voyageropenendedembodiedagent} introduced self-generated curricula to enable open-ended exploration in Minecraft, while Cradle leveraged multimodal LLMs to operate in visually rich, unstructured games such as Red Dead Redemption 2. Beyond game-specific agents, several efforts have explored generalist interaction with arbitrary user interfaces and environments, as well as the use of LLM agents to model social behaviors and multi-agent coordination in virtual worlds.\cite{park2023generativeagentsinteractivesimulacra, wang2025leveraging, qin2025uitarspioneeringautomatedgui}\\\\
\textbf{Reinforcement learning Assisted Gameplay Testing}. Several works proposed the use of Reinforcement Learning for automated game testing. \cite{8952543}, presented \textit{Wuji}, which aims to generate effective policies that explore more game states where bugs may be present, using evolutionary multi-objective optimization (EMOO) and deep reinforcement learning (DRL). Similarly, \citet{agarwal2020visualizing} used RL and an evolutionary algorithm to understand the navigation behavior in level design, with the main contribution being the web-based visualization tool. This work was focused on a specific 2D game called \textit{Sonic the Hedgehog 2}. Recent works have used exploration rewards for game testing, \cite{bergdahl2021augmentingautomatedgametesting} designed an environment with bugs. RL agent with reward signals to navigate to explore new states, showing an increase in test coverage. Similar to this work, \cite{gordillo2021improvingplaytestingcoveragecuriosity, 9970382, liu2022inspectorpixelbasedautomatedgame, Le_Pelletier_de_Woillemont_2022} used curiosity-based intrinsic reward for exploration, prioritizing state coverage for automated game-testing. Especially, \cite{9970382} proposed an approach to train agents capable of exploring while imitating an expert demonstration. EVOLUTE \cite{amadori2024robustimitationlearningautomated} is a recent approach that uses imitation learning by combining behavioral cloning with energy-based models to handle both discrete and continuous actions.\\\\
Previous work in automated game testing has largely focused on maximizing \emph{state coverage}~\cite{gordillo2021improvingplaytestingcoveragecuriosity, 9970382, liu2022inspectorpixelbasedautomatedgame}. Although this strategy can surface untested areas of the environment, it does not guarantee adequate coverage of the underlying codebase. To our knowledge, we are the first to incorporate call stack information into reinforcement learning for automated game testing.

\section{PRELIMINARIES}
\begin{figure*}[ht!]
\centering
\includegraphics[trim={1.0cm 0cm 1.0cm 0cm}, clip, width=0.72\linewidth]{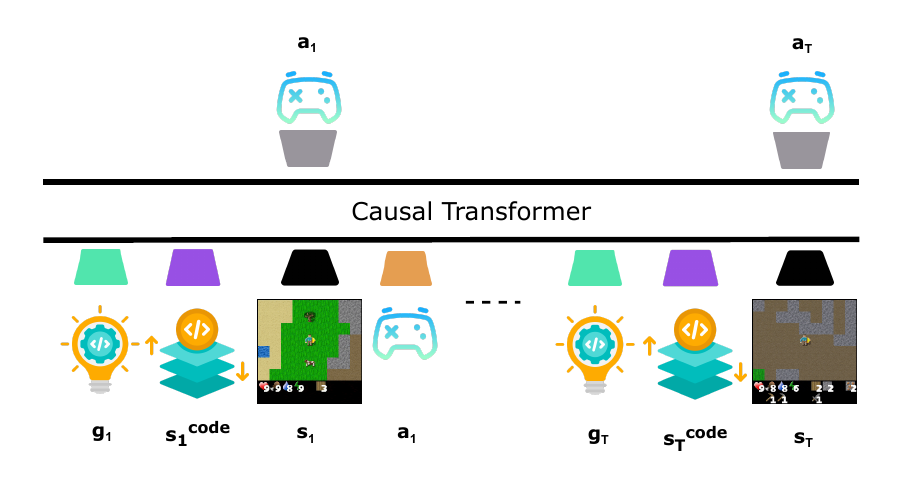}
\vspace{-5mm}
   \caption{Illustration of our Code-Aware Agent. All architectures use similar input formats, where a goal $g$, call stack $s^\mathsf{code}_t$, environment state $s_t$, and action $a_t$ are embedded and used as input.}
   \label{fig:pipeline}
\end{figure*}
\textbf{Problem Statement:} This is an offline learning approach, and the agent's goal is to learn and successfully reach the functions of interest from the existing data without the privilege of learning from online exploration. For this work, we only have access to a fixed, limited dataset consisting of a trajectory rolled out by game testers.\\\\

\textbf{Offline Reinforcement Learning:} A code-aware MDP is described by a tuple $(S, C, A, P, G)$ where $(S, A, P)$ are the state space, action space, and transition of the agent's MDP respectively, C is a finite set of code functions, and $G \subset C$ is a set of goal (target) functions. The code-aware MDP induces a new MDP $(S \times S^{\mathsf{code}}, A, P)$ where $S^{\mathsf{code}}$ is a space of call stacks. The agent in this MDP will be goal-conditioned, given a goal $g\in G$, it tries to reach a state $x = (s, s^\mathsf{code})$ such that $g \in s^{\mathsf{code}}$. The length of the number of function calls can be of arbitrary length, varying from one timestep to another based on the function calls. The trajectory $\tau$  is represented as follows $(s_0, s^\mathsf{code}_0, a_0, s_1, s^\mathsf{code}_1, a_1$ $\cdots s_T, s^\mathsf{code}_T, a_T)$, where $s^\mathsf{code}_t$ is a list of function call represented as $[f_t^1, f_t^2 \cdots f_t^{F}]$, here $f\in C$  and $F$ denotes the length of the call stack, $t\in [0, T]$ and $T$ is the length of the trajectory. \\\\
\textbf{Goal-Conditioned Behavioral Cloning (GCBC):}~\citep{lynch2019learninglatentplansplay} performs behavioral cloning using future states within the same trajectory as goals. In our setting, the goal $g$ corresponds to a target function name (from the call stack), and the agent learns a goal-conditioned policy $\pi(a \mid s, s^\mathsf{code}, g)$, where $s^\mathsf{code}$ represents the call stack at the current state.\\
The objective used to train the policy depends on the action space:
\begin{equation}
    \label{eq:gcbc_ce}
    J^{\mathsf{disc}}_{GCBC}(\pi) =
    \mathbb{E}_{(s,s^\mathsf{code},a,g)\sim\mathcal{D}}  \left[ \log \pi(a \mid s, s^\mathsf{code}, g) \right],
\end{equation}
\begin{equation}
\label{eq:gcbc_mse}
    J^{\mathsf{cont}}_{GCBC}(\pi) =
    \mathbb{E}_{(s,s^\mathsf{code},a, g)\sim\mathcal{D}}    \left[ \| \pi(s, s^\mathsf{code}, g) - a \|^2 \right],
\end{equation}
where $\mathcal{D}$ is a set of transition tuples $(s, s^\mathsf{code}, a, r, s', s^{\mathsf{code}'}, g)$, where $(s', s^\mathsf{code'})$ is a code-aware state visited upon taking action a from code-aware state $(s, \mathsf{code})$, and $g$ is a goal---Dataset construction is discussed in detail in Section~\ref{sec:exp}. Equations~\ref{eq:gcbc_ce} and~\ref{eq:gcbc_mse} correspond to the loss functions used in discrete and continuous action space, respectively.\\\\
\textbf{Goal-Conditioned Implicit Q-Learning (GCIQL)} Goal-conditioned variant of Implicit Q-Learning (IQL)~\cite{kostrikov2021offlinereinforcementlearningimplicit}, which learns value functions using expectile regression. In our setup, both the value and Q-functions are conditioned on the call stack $s^\mathsf{code}$ in addition to the state and goal. GCIQL learns $Q(s, s^\mathsf{code}, a, g)$ and $V(s, s^\mathsf{code}, g)$ by minimizing the following losses:
\begin{equation}
\begin{split}
    \mathcal{L}_V^{\text{GCIQL}}(V) &= \mathbb{E}_{(s,s^\mathsf{code},a, g)\sim\mathcal{D}} 
    \left[ \ell^2_\kappa\left( \bar{Q}(s, s^\mathsf{code}, a, g) - V(s, s^\mathsf{code}, g) \right) \right],
\end{split}
\end{equation}
\begin{equation}
\begin{split}
    \mathcal{L}_Q^{\text{GCIQL}}(Q) &= \mathbb{E} _{(s,s^\mathsf{code},a,s', s^\mathsf{code'},g)\sim\mathcal{D}} 
    \left[ \left( r(s, s^\mathsf{code}, g) + \gamma V(s', s^\mathsf{code'}, g) - Q(s, s^\mathsf{code}, a, g) \right)^2 \right], 
\end{split}
\end{equation}
where $r(s, s^\mathsf{code}, g) = \mathbb{I}\{g \in s^\mathsf{code}\} - 1$ is the sparse goal-conditioned reward function, $\bar{Q}$ is the target Q-function, and $\ell^2_\kappa$ is the expectile loss with expectile parameter $\kappa$, which is set to 0.8. To extract a policy from the learned value functions, we use a behavior-regularized actor update based on TD3+BC~\citep{fujimoto2021minimalistapproachofflinereinforcement}:
\begin{equation}
\begin{split}
    J_{\text{TD3+BC}}(\pi) & = \mathbb{E}_{(s,s^\mathsf{code},a,s', s^\mathsf{code'},g)\sim\mathcal{D}}  \left[  Q(s, s^\mathsf{code}, \pi(s, s^\mathsf{code}, g), g) - \lambda~ J_{GCBC}(\pi) \right],
\end{split}
\end{equation}
where $\lambda$ controls the trade-off between Q-value maximization and staying close to the behavior policy using BC loss.  \\\\

\textbf{Decision transformers:} The decision transformer (DT) architecture was proposed by  Lili et al. \cite{chen2021decision} to efficiently handle the sequence modeling problem. DT outputs the optimal actions by conditioning an autoregressive model on the desired
return (reward), past states, and action. The simplicity and scalability of the model allow us to encode multi-modal data. For our work, it's the observation and call stack. The \textbf{Transformer} architecture introduced by \cite{vaswani2023attentionneed}, are powerful model designed to handle sequential data efficiently. The core idea behind Transformers is the self-attention mechanism, which allows the model to focus on different parts of the input sequence when making predictions. A Transformer is built using multiple layers of self-attention and feedforward networks, each with residual connections to help with training. In each self-attention layer, the input sequence is first converted into a set of $n$ embeddings $\{x_i\}_{i=1}^n$, one for each token. These embeddings are then transformed into queries $q_i$, keys $k_i$, and values $v_i$ through learned linear projections. The output embedding $z_i$ for the $i$-th token is computed as a weighted sum of all value vectors, where the weights are determined by the similarity between the $i$-th query and each key:
\begin{equation}
z_i = \sum_{j=1}^{n} \text{softmax}\left( \left\{ \langle q_i, k_{j'} \rangle \right\}_{j'=1}^{n} \right)_j \cdot v_j ,
\end{equation}
This mechanism allows the model to learn which parts of the input are most relevant to each other, making Transformers especially effective for tasks involving long-range dependencies.

\section{METHOD}
\vspace{-1mm}
In this section, we describe our core contribution: augmenting goal-conditioned reinforcement learning agents with call stack information. We present 1) Call Stack encoding and 2) our adaptation of the Decision Transformer~\cite{chen2021decision} for automated gameplay testing. Unlike the original framework that conditions on scalar rewards or return-to-go, our formulation conditions the policy on a desired goal and incorporates call stack traces at each timestep. This setup shifts the focus from optimizing for cumulative rewards to achieving programmatic objectives, making it well-suited for test generation tasks. The overall architecture is illustrated in Figure~\ref{fig:pipeline}, and Algorithm~\ref{alg:gc_decision_transformer} highlights differences compared to the original Decision Transformer (DT) algorithm~\cite{chen2021decision}. 
\subsection{CALL STACK ENCODING}

At each timestep $t$, we include the call stack $s^\mathsf{code}_t$ as an ordered list of function calls. We experiment with three different approaches to encode $s^\mathsf{code}_t$, which consists of an ordered list of function $[f^1, f^2 \cdots f^{F}]$:
\begin{itemize}
    \item \textbf{Simple Token Embedding}: \label{sec:simple}Each function in $s^\mathsf{code}$ is embedded via an embedding table\footnote{A simple lookup table that stores embeddings of a fixed dictionary and size} $\mathcal{E} : C \to \mathbb{R}^d$, where $d$ denotes the embedding dimension. The resulting vectors are summed to form a single embedding for the call stack.
    \begin{equation}
        \text{CS}_{emb} = \sum_{i=1}^{F} \mathcal{E}(f^i)
    \end{equation}
    \item \textbf{Linear Projection}: \label{sec:linear}We obtain function embeddings from the embedding table \(\mathcal{E}\), apply a learned linear projection \(W\in \mathbb{R}^{d \times d}\) to each embedding in the sequence, and then sum the resulting vectors:
    \begin{equation}
        \text{CS}_{emb} = \sum_{i=1}^{F} W \cdot \mathcal{E}(f^i)
    \end{equation}
    \item \textbf{Multi-Head Self-Attention (MHSA)}: \label{sec:mhsa}Function embeddings are processed using a Multi-Head Self-Attention (MHSA) block \cite{vaswani2023attentionneed}. The resulting output tokens are then aggregated by summation into a single context vector:
    \begin{equation}
        \text{CS}_{\text{emb}} = \sum_{i=1}^{F} \text{MHSA}([\mathcal{E}(f^1), \mathcal{E}(f^2), \dots, \mathcal{E}(f^F)])_i
    \end{equation}

\end{itemize}
In the case of DT $\text{CS}_{emb}$ is concatenated with other token embeddings, as shown in Figure~\ref{fig:pipeline}. Whereas in GCBC and GCIQL, $\text{CS}_{emb}$ is concatenated along with the goal and the state representation. We find that using a transformer-based encoder for the call stack leads to improved generalization and executing desired test behaviors. 

\subsection{AGENT: GC-DT}

\begin{algorithm}[htb]
\caption{Goal Conditioned DT with Call Stack}
\label{alg:gc_decision_transformer}
\definecolor{codeblue}{rgb}{0.28125,0.46875,0.8125}
\lstset{
    basicstyle=\fontsize{8pt}{8pt}\ttfamily\bfseries,
    commentstyle=\fontsize{8pt}{8pt}\color{codeblue},
    keywordstyle=
}
\begin{lstlisting}[language=python]
# g, cs, s, a, t: goal, call stack, state, action, timestep
# transformer: GPT Style
# embed_g, embed_a: Linear Embedding Layers
# embed_s: Conv / Linear Embedding Layers
# encode_cs: encoder for call stack
# embed_t: learned timestep embedding
# pred_a: action prediction head

# main model
def CodeAwareAgent(g, cs, s, a, t):
  # shared timestep embedding
  pos_emb = embed_t(t)
  
  # compute embeddings
  g_emb = embed_g(g) +  pos_emb
  # one token per call stack
  cs_emb = encode_cs(cs) + pos_emb  
  s_emb = embed_s(s) + pos_emb
  a_emb = embed_a(a) + pos_emb
  
  # interleave input as 
  # (g, cs_1, s_1, a_1, ..., cs_K, s_K, a_K)
  input_emb = stack(g_emb, cs_emb, s_emb, a_emb)
  
  # pass through transformer
  hidden_states = transformer(input_emb=input_emb)
  
  # select action token embeddings for prediction
  a_hidden = unstack(hidden_states).actions
  
  # predict action
  return pred_a(a_hidden)
\end{lstlisting}
\end{algorithm}
Our goal is to model behavior that leads to the execution of specific functions, which are used as goals during training and inference. To support goal-conditioned generation, we represent trajectories using a sequence of tokenized inputs that include the desired goal $g$, the call stack $s^\mathsf{code}_t$ at each timestep, the environment state $s_t$, and the action $a_t$ taken.\\
This leads to the following autoregressive trajectory format:
\begin{equation}
    \tau = \left(g, s^\mathsf{code}_1, s_1, a_1, s^\mathsf{code}_2, s_2, a_2, \dots, s^\mathsf{code}_T, s_T, a_T\right),
\end{equation}
where $g$ is fixed for a given trajectory and is prepended to the sequence as conditioning. Unlike standard Decision Transformers that condition on returns-to-go, we explicitly provide a function-level goal $g$ that corresponds to a code path we want the agent to trigger. This enables the model to generate actions that steer the agent toward desired programmatic behavior, rather than achieving numerical rewards.\\\\
\textbf{Training and inference.}
During training, the model learns to predict the next action given the goal, call stack, and state:
\begin{equation}
    \pi(a_t \mid g, s^\mathsf{code}_{\leq t}, s_{\leq t}, a_{<t}).
\end{equation}
At test time, the agent is given a goal $g$ and the initial environment state. The model generates actions autoregressively, using the evolving call stack and state history as context. This allows the agent to repeatedly test whether specific functionality is covered under different gameplay conditions.\\\\
\textbf{Architecture.}
We feed the last $K$ timesteps into our Goal-Conditioned Decision Transformer, resulting in a sequence of $4K$ tokens: one for the goal, call stack, state, and action. Each input modality is embedded using a learned projection: we apply a linear layer to the state and action, and a Multi-Head Self-Attention encoder for the call stack. This is followed by layer normalization~\cite{ba2016layernormalization} to stabilize training. For environments with visual observations, such as crafter, the state is first processed through a convolutional encoder, which outputs a latent vector before projection into the token embedding space. Additionally, each timestep is assigned a learned timestep embedding, which is added to all tokens corresponding to that timestep. \\

\vspace{-6mm}
\section{ENVIRONMENT}
We built two environments that support RL training and code profiling. These environments are built on top of Godot \cite{beeching2021godotreinforcementlearningagents} and Crafter \cite{hafner2022benchmarkingspectrumagentcapabilities}. 

\subsection{GODOT}
\label{sec-godot}
Godot is an open-source game engine, and Godot RL Agent \cite{beeching2021godotreinforcementlearningagents} is customized for Deep Reinforcement Learning (DRL) research. It provides an interface between the game engine and the RL algorithm.  In Godot Multilevel Robot, the bot has to start from an initial location and must navigate to higher levels. To enable code-based information, we manually modified the game engine code to incorporate a code state (call stack), along with the usual observations. The agent receives feedback on the corresponding function calls with each action taken in the environment. The observation space is a $1$D vector that includes Raycast data relative to the player and the distance to collectibles. In addition to this, we also append the function calls triggered by the action that led to this observation. We identified $34$ gameplay-related functions that are called at each action step. The environment uses an action repeat of $8$. To reduce redundancy, we exclude functions that are executed every frame regardless of the agent's actions. After eliminating the unnecessary functions, we ended up with $24$ interesting functions that can be potential goals for the RL agents. 

\subsection{CRAFTER}
\label{sec-crafter}
Crafter is an open-world survival game with pixel-based observation that evaluates a wide range of general abilities within a single environment. There were a total of $293$ functions related to the gameplay that were being tracked for each action step. Similar to the filtering performed on Godot, we eliminated functions that were invoked at every action step, regardless of the agent's behavior. This left us with $94$ distinct functions. These functions represent meaningful gameplay events and can serve as potential goals for regression testing using RL agents. The state space consists of an RGB image and the function calls triggered by the action that led to this observation. We kept the instrumentation minimal to make it easy for others to use. Instead of adding changes that are only useful for game testing. Further details can be found in the Appendix. 

\subsection{DATA COLLECTION}
For this work, we collect data by manually playing the game to emulate the real game testing scenario, where we have a human tester collecting player data. For Crafter, we collected $300$ trajectories with approximately $100$k samples, and for Godot, which was a relatively simpler game compared to Crafter. We stopped at around $20$k samples, $200$ trajectories.

\section{EXPERIMENTS}
\label{sec:exp}
We divide our experiments into two main parts. The primary focus is the offline setting, where the agent leverages existing gameplay data collected from human testers. In this setting, we compare \ours{} against other goal-conditioned offline reinforcement learning methods to assess function coverage and gameplay performance. Next, we present results with different design choices, demonstrating that \okey~ (use of call stack information with Multi-Head Self Attention) significantly aids learning. We compare it with simple encoding techniques. Finally, we also conduct an ablation study to validate the effectiveness of our design choices that are meaningful to the agent’s performance.

\subsection{EVALUATION SETUP}

All reported values are averaged over $5$ random seeds, with the best-performing model per seed selected based on the maximum success rate. To ensure fairness, all algorithms are trained with the same number of gradient updates and evaluated under identical conditions. During Evaluation, scores are averaged across 10 episodes for each target function. In each episode, the agent is provided with a specific function-level goal and is evaluated on its ability to reach this goal. Importantly, the episode continues even after the goal is reached, allowing the agent to interact with the environment further. This enables comprehensive measurement of episodic return and function-level coverage. Function coverage is computed after all evaluation episodes have been completed. It reflects whether the agent has successfully triggered various functions at least once throughout the entire evaluation set, not necessarily in every episode. This metric captures the breadth of the agent's exploratory and functional behaviors, complementing the goal-specific success rate without being redundant with it. Notably, high function coverage may occur even when the success rate is lower, indicating that the agent is still visiting and exercising meaningful parts of the codebase, which is beneficial for robust testing.

\subsection{EVALUATION METRICS}
In our offline experiments, we evaluate agent performance using three key metrics:
\begin{enumerate}
    \item \textbf{Success Rate:} This is the primary metric, defined as the percentage of test-time episodes in which the agent successfully reaches the target function specified by the goal. An episode is considered successful if the agent reaches the desired function within the allowed trajectory length. For each goal, we run 10 episodes and report the average success rate.

    \item \textbf{Episode Return:} The cumulative reward collected during an episode. This reflects how effectively the agent performs within the environment, according to the task-specific reward structure.

    \item \textbf{Function Coverage:} The number of unique functions triggered by the agent during an evaluation episode. This metric captures how extensively the agent interacts with the game environment and provides insight into its behavioral diversity.
\end{enumerate}
Together, the success rate serves as a direct indicator of goal achievement, while function coverage and reward provide complementary insights into the agent’s behavior.

\section{RESULTS}
In this section, we compare the performance of \ours~ in offline settings across two game environments, using 3 different metrics. We will also try to understand the reason behind these results in detail.

\subsection{OFFLINE RESULTS}
The results in Table~\ref{tabOffExp} show that \ours~ significantly improves both task performance and function coverage compared to standard behavior cloning (BC) and goal-conditioned (GC) baselines. Specifically, GC agents equipped with code awareness consistently outperform their non-code-aware counterparts across all metrics. This highlights that incorporating structured program signals like the call stack helps the agent not only achieve a better success rate but also explore the functional space of the environment more effectively without losing the ability to play the game. Notably, the GC-DT + \okey~ model achieves the highest success rates and function coverage, indicating that it can both accomplish game objectives and visit meaningful states. This aligns well with the idea of regression testing, where the agent must test key functions to verify behavioral correctness, suggesting that our method is capable of understanding and replaying functional structure in the environment. Among the GC methods, GC-DT shows superior performance compared to GC-IQL and GC-BC. This improvement can be attributed to the use of a context window, which allows the model to reason over longer temporal horizons and condition more effectively on the target function. Overall, GC agents with code awareness demonstrate superior generalization and goal-following ability over simple BC and goal-agnostic baselines.

\begin{table}[ht!]
\centering
\resizebox{\columnwidth}{!}{%
\begin{tblr}{
  cell{1}{3} = {c=2}{c},
  cell{2}{3-8} = {c},
  cell{1}{8} = {c},
  cell{1}{9} = {c},
  cell{1}{5} = {c=2}{c},
  cell{1}{7} = {c=2}{c},
  cell{3}{1} = {r=2}{},
  cell{5}{1} = {r=2}{},
  cell{7}{1} = {r=2}{},
  vline{3} = {1}{0.2em},
  vline{3-4,6,8} = {1}{},
  vline{3,5,7,9} = {1-8}{0.2em},
  vline{ 4, 6, 8} = {2-8}{},
  hline{1,3,5,7,9} = {-}{0.2em},
  hline{2} = {3-9}{0.2em},
  hline{4,6,8} = {2-9}{},
}
                     &         & GC-BC  &      & GC-IQL &      & GC-DT  &      & BC   \\
                     &         & Simple & \okey  & Simple & \okey  & Simple & \okey  &      \\
{\small{Success} \\\small{Rate}}     & \small{Crafter} & \small{$39.67\pm0.52$}     & \textcolor{blue}{\small{$\mathbf{42.65\pm0.72}$}}  & \small{$41.78\pm0.60$}     & \textcolor{blue}{\small{$\mathbf{43.28\pm0.63}$}}   & \small{$48.78\pm0.35$}     & \textcolor{blue}{\small{$\mathbf{56.47\pm0.49}$}}   &  ~~~-    \\
                     & \small{Godot}   & \small{$25.45\pm0.29$}     & \textcolor{blue}{\small{$\mathbf{26.93\pm0.50}$}}   & \small{$23.88\pm0.52$}     & \textcolor{blue}{\small{$\mathbf{25.56\pm0.53}$}}   & \small{$41.22\pm0.54$} & \textcolor{blue}{\small{$\mathbf{47.15\pm0.77}$}}&  ~~~-   \\
{\small{Episode}\\\small{Return}}    & \small{Crafter} &\small{$1.62\pm0.16$}   & \textcolor{blue}{\small{$\mathbf{2.18\pm0.15}$}} & \small{$3.63\pm0.18$}    & \textcolor{blue}{\small{$\mathbf{4.08\pm0.19}$}}  & \small{$4.90\pm0.58$}    & \textcolor{blue}{\small{$\mathbf{5.12 \pm0.60}$}}  & \small{$1.65\pm0.31$} \\
                     & \small{Godot}   & \textcolor{blue}{\small{$\mathbf{8.17\pm0.39}$}}     & \small{$8.12\pm0.79$}   & \textcolor{blue}{\small{$\mathbf{8.13\pm0.37}$}}     & \small{$8.08\pm0.19$}   & \small{$13.12\pm0.59$}   & \textcolor{blue}{\small{$\mathbf{16.15\pm1.12}$}}   & \small{$7.90\pm1.51$}     \\
{\small{Function}\\\small{Coverage}} & \small{Crafter} & \small{$81.24\pm 0.42$}   & \textcolor{blue}{\small{$\mathbf{81.93\pm 0.37}$}} & \small{$80.74\pm 0.72$}   & \textcolor{blue}{\small{$\mathbf{84.20\pm 0.63}$}} & \small{$92.96\pm0.00$}   & \textcolor{blue}{\small{$\mathbf{96.92\pm0.00}$}} & \small{$79.44\pm 2.13$} \\
                     & \small{Godot}   & \textcolor{blue}{\small{$\mathbf{61.23\pm 0.47}$}}   & \small{$61.18\pm 0.68$} & \small{$60.97\pm 0.37$}   & \textcolor{blue}{\small{$\mathbf{61.86\pm 0.55}$}} & \small{$71.96\pm 0.24$}   & \textcolor{blue}{\small{$\mathbf{79.49\pm 1.88}$}} & \small{$60.10\pm 1.12$}    

\end{tblr}%
}
\caption{Performance of various offline agents across \textit{Crafter} and \textit{Godot}, evaluated using three metrics: Success Rate, Episode Return, and Function Coverage. ``Simple'' refers to non-code-aware agents, while ``\okey~'' denotes code-aware variant. Here, the BC agent is not conditioned on the goal function, hence the corresponding Success Rates are empty. 
\label{tabOffExp}}
\end{table}

\vspace{-1mm}
\subsection{EFFECT OF CONTEXT LENGTH}

We study the impact of context length in the decision transformer (Table~\ref{tab:ab_context}). This choice of $L$ was decided based on the episode length corresponding to each environment. Without \okey, performance saturates around a context of 10-20. With \okey, longer contexts (20–30) lead to better performance, particularly on Godot, where the success rate jumps from 30\% to 47\%. This shows that code-aware agents need more context to understand longer execution histories and make better decisions over time.
\begin{table}[ht!]
\resizebox{\columnwidth}{!}{%
\centering
\begin{tblr}{
  cell{1}{3} = {c=6}{c},
  cell{2}{3} = {c=2}{c},
  cell{2}{5} = {c=2}{c},
  cell{2}{7} = {c=2}{c},
  cell{4}{1} = {r=2}{},
  cell{3}{3} = {c},
  cell{3}{4} = {c},
  cell{3}{5} = {c},
  cell{3}{6} = {c},
  cell{3}{7} = {c},
  cell{3}{8} = {c},
  cell{4}{1} = {c},
  cell{5}{1} = {c},
  vline{3} = {-}{0.2em},
  vline{5,7} = {2-5}{0.2em},
  vline{4,6,8} = {3-5}{},
  hline{1,4,6} = {-}{0.2em},
  hline{2,3} = {3-8}{0.2em},
  hline{5} = {2-8}{},
}
 &        & GC-DT      &          &           &       &              &        \\
 &        & L          &          & 2L        &       & 3L            &         \\
 &        & Simple      & \okey   & Simple    & \okey  & Simple       & \okey    \\
 {\small{Success} \\\small{Rate}} & Crafter & $48.62,
\pm0.64$ & \textcolor{blue}{\textbf{$\mathbf{55.62\pm1.33}$}}   & $48.71\pm0.66$ & \textcolor{blue}{\textbf{$\mathbf{56.60\pm0.53}$}} & $48.78\pm0.35$ & \textcolor{blue}{\textbf{$\mathbf{56.47\pm0.49}$}} \\
 & Godot   & $26.67\pm0.95$ & \textcolor{blue}{\textbf{$\mathbf{30.90\pm0.76}$}} & $36.41\pm0.85$ & \textcolor{blue}{\textbf{$\mathbf{40.14\pm2.04}$}} & $41.22\pm0.88$ & \textcolor{blue}{$\mathbf{47.15\pm0.77}$}
\end{tblr}}
\caption{This table summarizes the performance of Decision Transformers across different context lengths. For Crafter, $L=10$, while for Godot $L=5$.  \label{tab:ab_context}}
\end{table}
\vspace{-6mm}
\subsection{DESIGN CHOICE}
Finally, we evaluate different architectural designs for integrating call stack information (Table~\ref{tab:ab-arch_design}) through an offline ablation study, using success rate as the primary metric. We find that a simple embedding of the call stack performs poorly. While adding a Multi-Head Self Attention (MHSA)\footnote{GC-DT + MHSA is our best \ours(\okey)} layer significantly improves performance, achieving the highest success rates on both Crafter (56\%) and Godot (47\%). These results highlight the importance of capturing dependencies between call stack elements, something MHSA is particularly well-suited for in our setting. 
\begin{table}[htb!]
\centering
\begin{tblr}{
  cell{1}{2} = {c=3}{c},
  cell{2}{3} = {c},
  cell{2}{4} = {c},
  cell{2}{5} = {c},
  vline{3, 4 } = {2-5}{},
  vline{2} = {-}{0.15em},
  hline{2-3} = {2-5}{},
  hline{2} = {2-5}{0.15em},
  hline{4} = {-}{},
  hline{1, 3, 5} = {1-5}{0.15em},
}
         & \small{GC-DT~(\okey)} &        &      \\
        & \small{Emb}   & \small{Linear} & \small{MHSA} \\
 \small{Crafter} & \small{$47.63\pm1.28$}    & \small{$49.15\pm0.87$}      & \textcolor{blue}{\small{$\mathbf{56.90\pm1.01}$}}   \\
 \small{Godot}   & \small{$34.53\pm1.46$}    & \small{$35.15\pm1.16$}      & \textcolor{blue}{\small{$\mathbf{39.40\pm6.78}$}}
\end{tblr}
\caption{This table summarizes the Success Rate of different Encoder choices. (Results averaged across all three context lengths)\label{tab:ab-arch_design}}
\end{table}
\vspace{-6mm}

\section{CONCLUSION}

In this work, we show that our Causal Auto-regressive Architecture improves agent performance, particularly when combined with longer context and self-attention. Our results demonstrate that incorporating code-level signals, such as call-stack information, helps agents better understand how their actions relate to the game logic, leading to more structured and effective testing.
\\\\
More broadly, this approach moves toward automated, code-aware game testing that reduces reliance on manual playtesting. While scalability remains a challenge due to the large size of modern game codebases, a modular testing strategy—where agents focus on specific features or submodules can make the system more practical. Integrating code-aware agents early in development could enable faster iteration, targeted debugging, and more reliable validation of game components.

\section{LIMITATION AND FUTURE WORK}
A key limitation of our approach is the high computational cost of collecting real-time function-level code coverage and call stack information in complex game environments. This instrumentation introduces runtime overhead and limits scalability to larger games with many functions and events. Moreover, treating all functions equally may not reflect practical developer priorities. In future work, we will focus on smaller, modular environments and explore selective instrumentation that tracks only developer-specified or high-priority functions. This would reduce overhead, improve practical relevance, and make the framework more scalable for real-world game development.

\section*{ACKNOWLEDGMENT}
We would like to thank Sarra Habchi, Ian Guak, Alessandro Palmas, and Gabriel Robert for their valuable discussions and feedback throughout this project. We also thank Harley Wiltzer and Mahtab (Mattie) Nejati for their helpful comments on the manuscript. This work was supported by the Mitacs PhD Accelerate Grant. This research was enabled in part by compute resources, software, and technical support provided by Ubisoft (La Forge).

\printbibliography[heading=subbibintoc]
\appendix
\clearpage
\section{\okey~ADDITIONAL DETAILS}
\subsection{HYPERPARAMETERS}
We present the hyperparameters (Table~\ref{tab:dt_hyperparams}) used for training our Decision Transformer with Code-Aware Agent (GC-DT-\okey) agent. This configuration reflects design choices tailored to effectively capture temporal dependencies, program structure, and goal representations in structured game environments. We detail model architecture parameters such as embedding dimensions, transformer depth, and attention heads, along with training settings including optimizer parameters, learning rate scheduler, and dropout.

\subsection{ENVIRONMENT DETAILS}
\begin{table}[hbt!]
\centering
\footnotesize
\caption{Environment Details for Crafter and Godot\label{tab:env_details}}
\begin{tabular}{lcc}
\toprule
\textbf{Attribute} & \textbf{Crafter} & \textbf{Godot} \\
\midrule
State Type & Image & Vector \\
State Dimension ($s$) & $64 \times 64 \times 3$ & $65 \times 1$ \\
Action Space & Discrete [0, 16] & Continuous [-1,1] \\
Action Dimension ($a$) & 17 & 2 \\
Number of Functions ($f$) & 293 & 34 \\
Number of Goals ($g$) & 94 & 24 \\
Code Trace ($s^{\mathsf{code}}$) & $[f^1, f^2, \ldots, f^F]$ & $[f^1, f^2, \ldots, f^F]$ \\
Function calls per step ($F$) & $[0, 600]$ & $[0,15]$ \\
\bottomrule
\end{tabular}
\end{table}
\begin{table}
\centering
\footnotesize
\caption{Decision Transformer Hyperparameters}
\label{tab:dt_hyperparams}
\begin{tabular}{ll}
\toprule
\textbf{Parameter} & \textbf{Value / Description} \\
\midrule
\multicolumn{2}{c}{\textit{General Settings}} \\
\midrule
Agent Type & \texttt{Simple} / \texttt{cs} \\
Batch Size & 128 \\
Epochs & 10 \\
Training Steps & 10000 \\
Evaluation Episodes & 10 \\
\midrule
\multicolumn{2}{c}{\textit{GPT Model Configuration}} \\
\midrule
Context Length & L / 2L / 3L \\
Value of $L$ & 5 (Godot) / 10 (Crafter) \\
Token Dimension & 128 \\
Transformer Layers ($n\_layer$) & 2 \\
Attention Heads ($n\_head$) & 2 \\
Dropout & 0.1 \\
Learning Rate & $3 \times 10^{-4}$ \\
Adam Betas & (0.9, 0.95) \\
Gradient Norm Clip & 1.0 \\
Weight Decay & 0.1 \\
LR Scheduler & Linear warmup + cosine decay \\
\midrule
\multicolumn{2}{c}{\textit{Image Encoder~(Crafter)}} \\
\midrule
Encoder Channels & 32, 64, 64 \\
Encoder Filter Sizes & $8 \times 8$, $4 \times 4$, $3 \times 3$ \\
Encoder Strides & 4, 2, 1 \\
\midrule
\multicolumn{2}{c}{\textit{State Encoder~(Godot)}} \\
\midrule
Number of Linear Layer & 2 \\
Hidden Dimension & 1024 \\
\midrule
\multicolumn{2}{c}{\textit{Call Stack Encoder:~(MHSA)}~\ref{sec:mhsa}} \\
\midrule
Function Embedding Dimension & 128 \\
Goal Embedding Dimension & 128 \\
Number of Layers & $1$ \\
Number of heads & $2$ \\
\bottomrule
\end{tabular}
\end{table}
Table~\ref{tab:env_details} provides environment specific details. The ``Function calls per step''$F$ indicates the total number of functions triggered by a single action. In Crafter, this number is relatively high, especially at the start of the game due to initialization, though it typically comes down around 100–200 function calls per step on average. In contrast, the number is significantly lower in Godot, primarily due to limited access to instrumentation within the Godot engine codebase.


We present the hyperparameters (Table~\ref{tab:dt_hyperparams}) used for training our Decision Transformer with Code-Aware Agent (GC-DT-\okey) agent. This configuration reflects design choices tailored to effectively capture temporal dependencies, program structure, and goal representations in structured game environments. We detail model architecture parameters such as embedding dimensions, transformer depth, and attention heads, along with training settings including optimizer parameters, learning rate scheduler, and dropout. 

\subsection{LATENCY}
Our implementation uses lightweight instrumentation that logs active function identifiers directly from the engine’s call stack interface, adding minimal overhead. We are currently profiling the Environment latency time per environment step with and without code instrumentation. Experiments show that instrumentation adds less than 5–8\% overhead in both Crafter and Godot environments, negligible relative to the total simulation time, suggesting that the overhead is dominated by environment simulation rather than by the instrumentation itself.

\subsection{COMPUTE REQUIREMENTS}
 All experiments were conducted on a machine equipped with an NVIDIA A4000 GPU and 16 GB of memory. The peak memory usage for the \okey~ agent was approximately $\sim 13.6GB$, while the simple agent required around $\sim4GB$.
\subsection{CODE}
The code for the \okey~ agent will be made publicly available on acceptance.


\end{document}